\documentclass{aa}  

\usepackage{graphicx}
\usepackage{lscape}
\usepackage{txfonts}
\begin{document}

   \title{On the Period--Luminosity--Metallicity relation of Classical Cepheids \thanks{Based on observations made with the Italian Telescopio Nazionale Galileo (TNG) operated by the Fundación Galileo Galilei (FGG) of the Istituto Nazionale di Astrofisica (INAF) at the Observatorio del Roque de los Muchachos (La Palma, Canary Islands, Spain).}}


\author{V. Ripepi \inst{1} 
          \and 
        G. Catanzaro \inst{2}
          \and 
          R. Molinaro \inst{1}
          \and
          M. Marconi \inst{1}
          \and 
          G. Clementini \inst{3}
          \and
          F. Cusano \inst{3}
          \and \\
          G. De Somma \inst{1,4,5}
          \and
          S. Leccia \inst{1}
          \and
          I. Musella \inst{1}
          \and
          V. Testa \inst{6}
}

\institute{ INAF-Osservatorio Astronomico di Capodimonte, Salita Moiariello 16, 80131, Naples, Italy\\  \email{vincenzo.ripepi@inaf.it}
\and 
INAF-Osservatorio Astrofisico di Catania, Via S.Sofia 78, 95123, Catania, Italy \\
             \email{giovanni.catanzaro@inaf.it}
             \and
INAF-Osservatorio di Astrofisica e Scienza dello Spazio, Via Gobetti 93/3, I-40129 Bologna, Italy 
             \and
Dipartimento di Fisica "E. Pancini", Università di Napoli "Federico II", Via Cinthia, 80126 Napoli, Italy
\and
Istituto Nazionale di Fisica Nucleare (INFN)-Sez. di Napoli, Via Cinthia, 80126 Napoli, Italy     \and        
             INAF – Osservatorio Astronomico di Roma, via Frascati 33, I-00078 Monte Porzio Catone, Italy
             }

   \date{}

 
  \abstract
   { Classical Cepheids (DCEPs) are the most important primary indicators for the extragalactic distance scale. Establishing the dependence on metallicity of their period--luminosity and period--Wesenheit (PL/PW) relations has deep consequences on the estimate of the Hubble constant (H$_0$).}   
   {We aim at investigating the dependence on metal abundance ([Fe/H]) of the PL/PW relations for Galactic DCEPs.}
   {We combined proprietary and literature photometric and spectroscopic data, gathering a total sample of 413 Galactic DCEPs (372 fundamental mode -- DCEP\_F and 41 first overtone -- DCEP\_1O) and constructed new metallicity-dependent PL/PW relations in the near infra-red (NIR) adopting the Astrometric Based Luminosity.}
   {
   We find indications that the slopes of the PL$(K_s)$ and PW$(J,K_S)$ relations for Galactic DCEPs might depend on metallicity when compared to the Large Magellanic Cloud relationships. Therefore, we have used a generalized form of the PL/PW relations to simultaneously take into account the metallicity dependence of the slope and intercept of these relations.
   }
   {We calculated PL/PW relations which, for the first time,  explicitly  include  a metallicity dependence of both the slope and intercept terms. Although  the  insufficient quality of the available  data  makes our results not yet  conclusive,  they are  relevant from a methodological point of view. 
  The new   relations are linked to the geometric measurement of the distance to the Large Magellanic Cloud and 
   allowed us to  estimate a {\it Gaia} DR2 parallax zero point  offset $\Delta \varpi$=0.0615$\pm$0.004 mas from the dataset of DCEPs used in this work.}

   \keywords{Stars: distances –- Stars: variables: Cepheids –- Distance scale -- Stars: abundances -- Stars: fundamental parameters}

   \maketitle
%

\section{Introduction}

Classical Cepheids (DCEPs) are the most important primary distance indicators of the cosmic distance 
ladder thanks to their Period-Luminosity \citep[PL,][]{Leavitt1912} and Period-Wesenheit
\citep[PW,][]{Madore1982} relations. Indeed, once calibrated locally by means of geometric methods such as trigonometric parallaxes, the PL/PW relations  can be used to calibrate secondary distance indicators such as Type Ia Supernovae (SNe), which are sufficiently powerful to measure distances in the unperturbed Hubble flow. Eventually, this three-step procedure allows us to 
 measure the Hubble constant (H$_0$) from the slope of the relation between the distance to far away galaxies in the Hubble flow and their recession velocity (e.g. \citealt{Sandage2006,Freedman2011,Freedman2012,Riess2016,Riess2019}).
 
 The value of H$_0$ is  presently a matter of vivid debate in the literature because the most recent determinations 
 based on the cosmic distance ladder,  H$_0$=74.03$\pm$1.42 km s$^{-1}$ Mpc$^{-1}$ (\citealt{Riess2019} and references therein) significantly disagree with the value of H$_0$ estimated from the Planck Cosmic Microwave Background (CMB) measurements under the flat $\Lambda$ Cold Dark Matter ($\Lambda$CDM) model,   H$_0$=67.4$\pm$0.5 km s$^{-1}$ Mpc$^{-1}$ \citep{Planck2018}. This 4.4$\sigma$ discrepancy, commonly referred to as the H$_0$ tension, remains  unexplained  notwithstanding several attempts to reduce the systematic errors affecting the different steps of the cosmic distance ladder  \citep[e.g.][]{Huang2020,Yuan2019,Reid2019}.  
Recently, a value of H$_0 = 73.3^{+1.7}_{-1.8}$ km s$^{-1}$ Mpc$^{-1}$, in agreement with the cosmic distance ladder, was obtained by the H0LICOW and STRIDES collaborations based on the gravitational time-delays of six lensed quasars \citep[][]{Wong2020}. Since this measurement is totally independent of the cosmic distance ladder calibrations, the occurrence of a tension appears to be confirmed.
In order to improve the accuracy of the H$_0$ measurements and to put stringent constraints on  cosmological models, it is crucial to quantify the residual systematic errors affecting  different methods to estimate H$_0$.  In particular, it is mandatory to quantify and reduce the uncertainties (random and systematic) of the different steps involved in the distance ladder calibration.

One of the main residual source of uncertainty in the cosmic distance ladder is the metallicity dependence of the DCEP PL/PW relations. Indeed, the metallicity is widely expected to affect both the slope and intercept of the DCEP $PL/PW$ relations, 
particularly at  
 optical passbands. Conversely, the metallicity dependence of the DCEP PW relations is estimated and predicted to be small for different band combinations but especially in the near infra-red  \citep[NIR, see e.g.][and references therein]{Fiorentino2007,Ngeow2012b,Dicriscienzo2013,Fiorentino2013,Gieren2018}. 
The procedure adopted by the SH0ES team  \citep{Riess2016} to calibrate the cosmic distance ladder, takes into account the metallicity dependence of the PW relations. According to these authors, the metallicity contributes a 0.5\% to the total error budget of 2.4\% in the measure of H$_0$. However, there is a significant  disagreement among estimates for the metallicity dependence of the PL/PW relations obtained by different authors  \citep[e.g.][]{Macri2006,Romaniello2008,Bono2010,Freedman2011,Shappee2011,Pejcha2012,Groenewegen2013,Kodric2013,Fausnaugh2015,Riess2016}. This variety of results can be partially explained by the significant errors still affecting the metal abundances of DCEPs in the distant galaxies which are addressed in these studies. 

A direct determination of the metallicity effect on the Cepheid distance scale, based on Galactic DCEPs with [Fe/H] abundances obtained from  high-resolution spectroscopy, has been  hampered until recently  by the lack of precise distance measurements for a sufficiently large number of MW DCEPs.
In this context, a unique  contribution is nowadays being provided by the {\it Gaia} mission \citep{Gaia2016} which published in its Data Release 2 (DR2)  precise parallaxes for hundreds of MW DCEPs, including most of the new  discoveries in the last couple of years \citep[see e.g.][]{Chen2018,Jayasinghe2018,Udalski2018,Clementini2019,Ripepi2019}. Further improved  parallaxes are foreseen to be provided by future {\it Gaia} data releases. 
\begin{figure*}
   \centering
   \includegraphics[width=10cm]{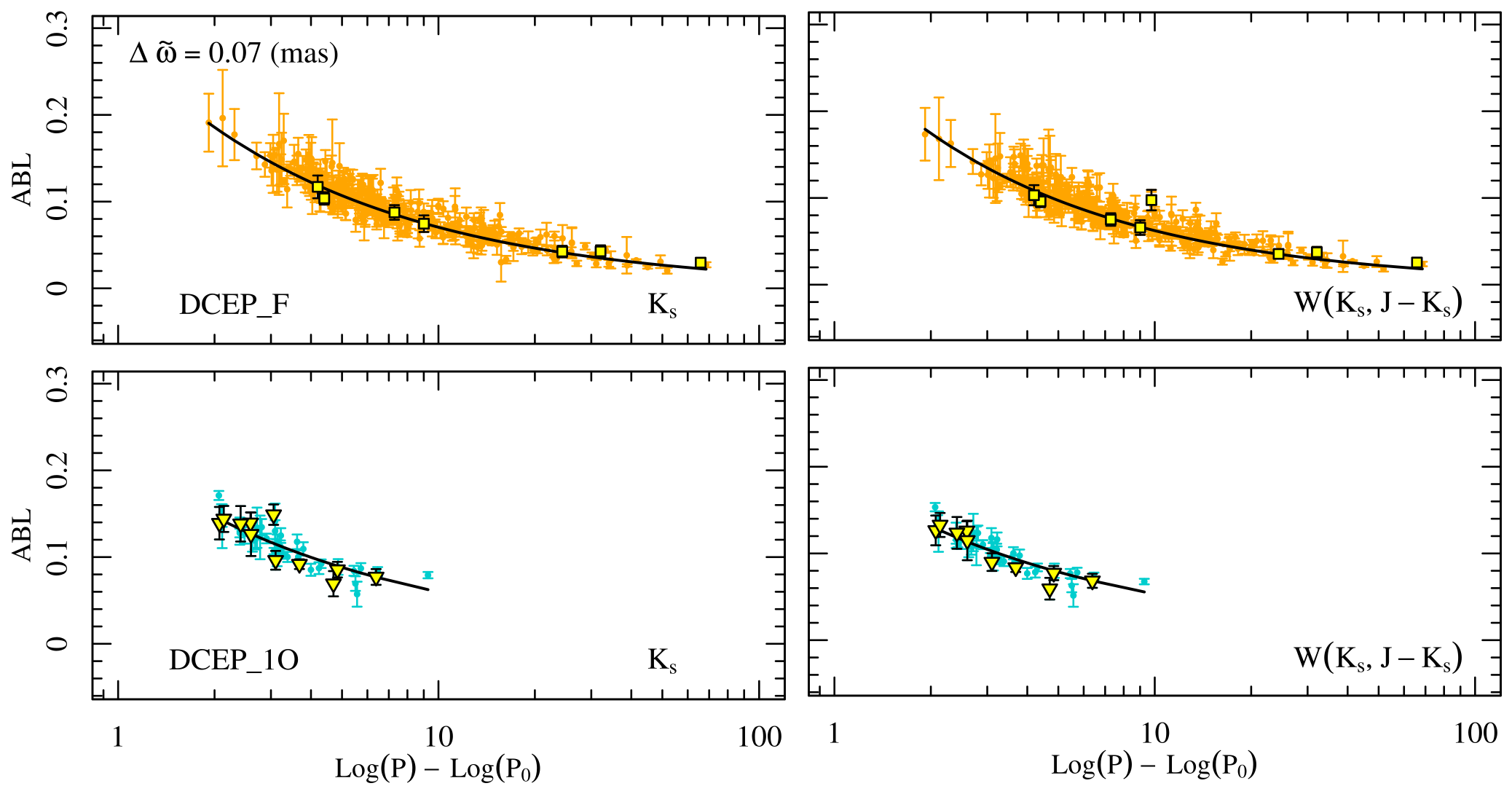}
   \caption{Results of the fitting procedure for an ABL in the form of Eq.~\ref{eqABL} (left) and Eq.~\ref{eqABL1} (right) for DCEP\_Fs (top panels) and DCEP\_1Os (bottom panels) and a {\it Gaia} zero point parallax offset of $\Delta \varpi=0.07$ mas.
  Orange and cyan filled circles represent the literature data for F and 1O MW DCEPs,  
   respectively. Yellow-filled squares and triangles display the DCEP\_Fs and DCEP\_1Os analysed in Paper I, respectively. For clarity reasons,  outliers were not shown.} 
   \label{fig:PW0}
    \end{figure*}

\begin{figure*}
   \centering
   \includegraphics[width=10cm]{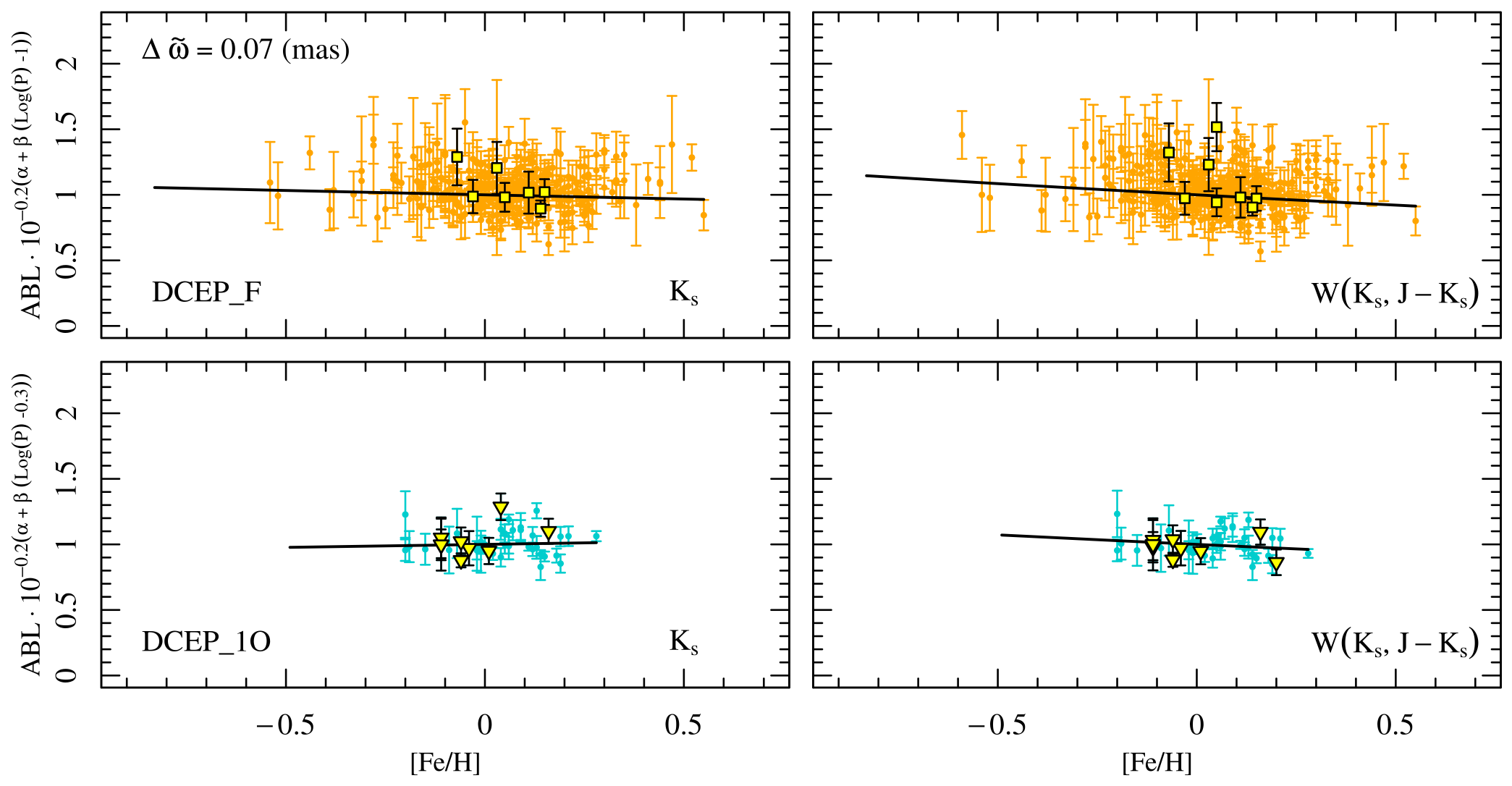}
   \caption{Results of the fitting procedure for an ABL in the form of Eq.~\ref{eqABLZ1} (left) and Eq.~\ref{eqABLZ2} (right) for DCEP\_Fs (top panels) and DCEP\_1Os (bottom panels) and a {\it Gaia} zero point parallax offset of $\Delta \varpi=0.07$ mas. Symbols are as in Fig.~\ref{fig:PW0}
  } 
   \label{fig:PW}
    \end{figure*}

However, only a small fraction of the recently discovered DCEPs  have accurate multiband photometry and spectroscopic abundances. 
Therefore, to fully  exploit the potential of such an  ideal amount of data and better estimate the metallicity dependence of the DCEP PL/PW relations,  
it is crucial to increase the sample of well characterized (both in  photometry and spectroscopy) MW DCEPs. 
To this end, we are carrying out a project which foresees the combination of multiband photometry availble in the literature with metal abundances from newly acquired  high-resolution spectroscopy for more than a hundred MW DCEPs, selected among the new discoveries and  those  poorly characterized in the literature so far. 
First results from this project were published in \citet{Catanzaro2020}, where we present a new lithium-rich DCEP, V363 Cas, {the fifth of such a rare class of DCEPs to be discovered in the MW}, and in Ripepi et al. (A\&A submitted, hereinafter Paper I) where we present elemental abundances from high-resolution spectroscopy for a first sample of 21 DCEPs and a Type II Cepheid of BL Herculis (BL Her) type {among the more than 100 DCEPs we are targeting with  our observing program.} 

In this paper we exploit the iron abundances published in Paper I along with literature data to study the metallicity dependence of the NIR PL/PW relations for MW DCEPs.   

%

\section{The sample}

In Paper I we presented chemical abundances and photometric properties for a sample of 21 DCEPs and one BL Her star\footnote{Star ASAS J162326-0941.0, was previously classified as DCEP.}. This sample was complemented with homogeneous literature data from \citet{Clementini2017} and \citet{Groenewegen2018}. A detailed description of the properties of our total sample can be found in Paper I. Here we only recall that 
it consists of 389 fundamental mode (DCEP\_F) and 44 first overtone (DCEP\_1O) classical Cepheids. Among them, those with RUWE$\geq 1.4$ were discarded as recommended by the {\it Gaia} documentation\footnote{The RUWE parameter measures the reliability of the {\it Gaia} parallaxes, see  Section 14.1.2 of "Gaia Data Release 2 Documentation release 1.2"; https://gea.esac.esa.int/archive/documentation/GDR2/.}. Similarly, we removed the DCEPs with  $V<6$ mag, as their {\it Gaia} parallaxes are too uncertain \citep[see][]{Riess2018}. The literature sample hence reduced to 364 DCEP\_F and 38 DCEP\_1O pulsators. Adding the 9 DCEP\_F  and 12 DCEP\_1O pulsators studied in Paper I,  our final sample consisting of 373 DCEP\_Fs and 50 DCEP\_1Os. 

\begin{table*}
\caption{Results from the least square fit of an ABL function in the
  form of Eqs.~\ref{eqABL} and ~\ref{eqABL1} (upper portion of the
  table);  Eqs.~\ref{eqABLZ1} and ~\ref{eqABLZ2} (middle portion of the table), or Eqs.~\ref{eqABLZbis1} and ~\ref{eqABLZbis2} (lower part of the table). The functional form of the PW relation is 
labelled in the table. $P_0$ is the pivoting period, whose logarithm is equal to 1.0 and 0.3 
 for DCEP\_Fs and DCEP\_1Os, respectively.  
 }             
\label{Tab:PW}      
\centering                          
\setlength{\tabcolsep}{3.5pt}
\begin{tabular}{ l c c c c r c l c c c}        
\hline\hline                 
\noalign{\smallskip} 
\multicolumn{1}{c}{$\alpha$ } &
\multicolumn{1}{c}{$\beta$ } &
\multicolumn{1}{c}{$\gamma$} &
\multicolumn{1}{c}{$\delta$} &
\multicolumn{1}{c}{n} &
\multicolumn{1}{c}{n$_{rej}$} &
\multicolumn{1}{c}{Mode} &
\multicolumn{1}{c}{$\Delta \varpi$} &
\multicolumn{1}{c}{Case} &
\multicolumn{1}{c}{$\sigma_{\rm ABL}$} &
\multicolumn{1}{c}{AIC}                           
\\
            \noalign{\smallskip}
\multicolumn{1}{c}{(1)} &
  \multicolumn{1}{c}{(2)} &
  \multicolumn{1}{c}{(3)} &
  \multicolumn{1}{c}{(4)} &
  \multicolumn{1}{c}{(5)} &
  \multicolumn{1}{c}{(6)} &
\multicolumn{1}{c}{(7)} &
\multicolumn{1}{c}{(8)} &
\multicolumn{1}{c}{(9)} &
\multicolumn{1}{c}{(10)} &
\multicolumn{1}{c}{(11)} 
\\
            \noalign{\smallskip}
\hline
            \noalign{\medskip}
\multicolumn{11}{c}{$M_{K_{S,0}}$ or $W(J,K_S)=\alpha+\beta(\log P-\log P_0)$}\\           
            \noalign{\medskip}
\hline
            \noalign{\smallskip}
$-5.840 \pm 0.023$ &  $-3.061 \pm 0.070$ &  -- & -- & 373 & 47 & F & 0.0490 & (K) & 0.0095 & $-$2594 \\ 
$-5.794 \pm 0.021$ &  $-3.027 \pm 0.090$ &  -- & -- & 373 & 55 & F & 0.0615 & (K) & 0.0091 & $-$2533 \\ 
$-5.762 \pm 0.028$ &  $-3.012 \pm 0.086$ &  -- & -- & 373 & 53 & F & 0.0700 & (K) & 0.0098 & $-$2533 \\ 
$-6.131 \pm 0.027$ &  $-3.219 \pm 0.078$ &  -- & -- & 373 & 45 & F & 0.0490 & (K, JK) & 0.0091 & $-$2666 \\ 
$-6.072 \pm 0.024$ &  $-3.185 \pm 0.085$ &  -- & -- & 373 & 51 & F & 0.0615 & (K, JK) & 0.0086 & $-$2635 \\ 
$-6.027 \pm 0.016$ &  $-3.200 \pm 0.087$ &  -- & -- & 373 & 40 & F & 0.0700 & (K, JK) & 0.0090 & $-$2631 \\ 
$-4.223 \pm 0.082$ &  $-2.836 \pm 0.474$ &  -- & -- & 50 & 2 & 1O & 0.0490 & (K) & 0.0109 & $-$366 \\ 
$-4.190 \pm 0.117$ &  $-2.788 \pm 0.533$ &  -- & -- & 50 & 2 & 1O & 0.0615 & (K) & 0.0101 & $-$364 \\ 
$-4.166 \pm 0.091$ &  $-2.780 \pm 0.453$ &  -- & -- & 50 & 2 & 1O & 0.0700 & (K) & 0.0107 & $-$365 \\ 
$-4.457 \pm 0.101$ &  $-2.889 \pm 0.425$ &  -- & -- & 50 & 2 & 1O & 0.0490 & (K, JK) & 0.0091 & $-$382 \\ 
$-4.430 \pm 0.092$ &  $-2.786 \pm 0.371$ &  -- & -- & 50 & 5 & 1O & 0.0615 & (K, JK) & 0.0078 & $-$363 \\ 
$-4.404 \pm 0.082$ &  $-2.802 \pm 0.428$ &  -- & -- & 50 & 4 & 1O & 0.0700 & (K, JK) & 0.0087 & $-$370 \\ 

  \noalign{\smallskip}
\hline
            \noalign{\medskip}
\multicolumn{11}{c}{$M_{K_{S,0}}$ or $W(J,K_S)=\alpha+\beta(\log P-\log P_0)+\gamma{\rm [Fe/H]}$}\\           
            \noalign{\medskip}
\hline
            \noalign{\smallskip}
$-5.837 \pm 0.030$ &  $-3.053 \pm 0.093$ &  $-0.039 \pm 0.151$ & -- & 373 & 46 & F & 0.0490 & (K) & 0.0096 & $-$2600 \\ 
$-5.785 \pm 0.027$ &  $-3.015 \pm 0.089$ &  $-0.082 \pm 0.138$ & -- & 373 & 55 & F & 0.0615 & (K) & 0.0091 & $-$2532 \\ 
$-5.746 \pm 0.022$ &  $-2.992 \pm 0.077$ &  $-0.141 \pm 0.140$ & -- & 373 & 54 & F & 0.0700 & (K) & 0.0098 & $-$2527 \\ 
$-6.121 \pm 0.027$ &  $-3.216 \pm 0.067$ &  $-0.084 \pm 0.145$ & -- & 373 & 41 & F & 0.0490 & (K, JK) & 0.0093 & $-$2684 \\ 
$-6.033 \pm 0.029$ &  $-3.171 \pm 0.065$ &  $-0.284 \pm 0.115$ & -- & 373 & 34 & F & 0.0615 & (K, JK) & 0.0093 & $-$2698 \\ 
$-5.991 \pm 0.027$ &  $-3.148 \pm 0.069$ &  $-0.356 \pm 0.171$ & -- & 373 & 33 & F & 0.0700 & (K, JK) & 0.0098 & $-$2686 \\ 
$-4.237 \pm 0.128$ &  $-2.869 \pm 0.641$ &  $0.225 \pm 0.255$ & -- & 50 & 1 & 1O & 0.0490 & (K) & 0.0108 & $-$368 \\ 
$-4.204 \pm 0.106$ &  $-2.787 \pm 0.426$ &  $0.161 \pm 0.295$ & -- & 50 & 3 & 1O & 0.0615 & (K) & 0.0098 & $-$357 \\ 
$-4.175 \pm 0.107$ &  $-2.739 \pm 0.517$ &  $0.101 \pm 0.298$ & -- & 50 & 5 & 1O & 0.0700 & (K) & 0.0088 & $-$341 \\ 
$-4.447 \pm 0.120$ &  $-2.893 \pm 0.412$ &  $-0.111 \pm 0.228$ & -- & 50 & 2 & 1O & 0.0490 & (K, JK) & 0.0101 & $-$381 \\ 
$-4.408 \pm 0.108$ &  $-2.788 \pm 0.427$ &  $-0.253 \pm 0.230$ & -- & 50 & 5 & 1O & 0.0615 & (K, JK) & 0.0082 & $-$363 \\ 
$-4.381 \pm 0.099$ &  $-2.761 \pm 0.484$ &  $-0.304 \pm 0.254$ & -- & 50 & 5 & 1O & 0.0700 & (K, JK) & 0.0079 & $-$362 \\ 
            \noalign{\smallskip}
\hline
            \noalign{\medskip}
\multicolumn{11}{c}{$M_{K_{S,0}}$ or $W(J,K_S)=\alpha+(\beta+\delta{\rm [Fe/H]})(\log P- \log P_0)+\gamma{\rm [Fe/H]}$} \\           
            \noalign{\medskip}
\hline
            \noalign{\smallskip}
$-5.837 \pm 0.023$ &  $-3.064 \pm 0.101$ &  $-0.044 \pm 0.170$ & $0.086 \pm 0.673$ & 373 & 49 & F & 0.0490 & (K) & 0.0096 & $-$2579 \\ 
$-5.771 \pm 0.030$ &  $-3.112 \pm 0.111$ &  $-0.185 \pm 0.191$ & $0.704 \pm 0.650$ & 373 & 30 & F & 0.0615 & (K) & 0.0111 & $-$2626 \\ 
$-5.731 \pm 0.033$ &  $-3.111 \pm 0.095$ &  $-0.256 \pm 0.158$ & $0.866 \pm 0.587$ & 373 & 27 & F & 0.0700 & (K) & 0.0113 & $-$2607 \\ 
$-6.130 \pm 0.035$ &  $-3.274 \pm 0.117$ &  $-0.078 \pm 0.164$ & $0.507 \pm 0.528$ & 373 & 38 & F & 0.0490 & (K, JK) & 0.0096 & $-$2690 \\ 
$-6.044 \pm 0.025$ &  $-3.234 \pm 0.118$ &  $-0.237 \pm 0.157$ & $0.465 \pm 0.592$ & 373 & 35 & F & 0.0615 & (K, JK) & 0.0096 & $-$2698 \\ 
$-6.005 \pm 0.025$ &  $-3.241 \pm 0.093$ &  $-0.283 \pm 0.160$ & $0.727 \pm 0.564$ & 373 & 26 & F & 0.0700 & (K, JK) & 0.0103 & $-$2716 \\ 
   \noalign{\smallskip}
\hline 
\noalign{\smallskip}
\end{tabular}
\tablefoot{The different columns provide: (1--4) coefficients of the non-linear fit
 and their relative errors; (5--6) total number of available DCEPs and
 number of sources  rejected during the fitting procedure; (7)
 pulsation mode; (8) adopted {\it Gaia} parallax offset (in mas); (9)
 Case; (10) r.m.s. of the residuals of the ABL function; (11) AIC
 value (see text).}
\end{table*}

\begin{figure*}
   \centering
   \includegraphics[width=10cm]{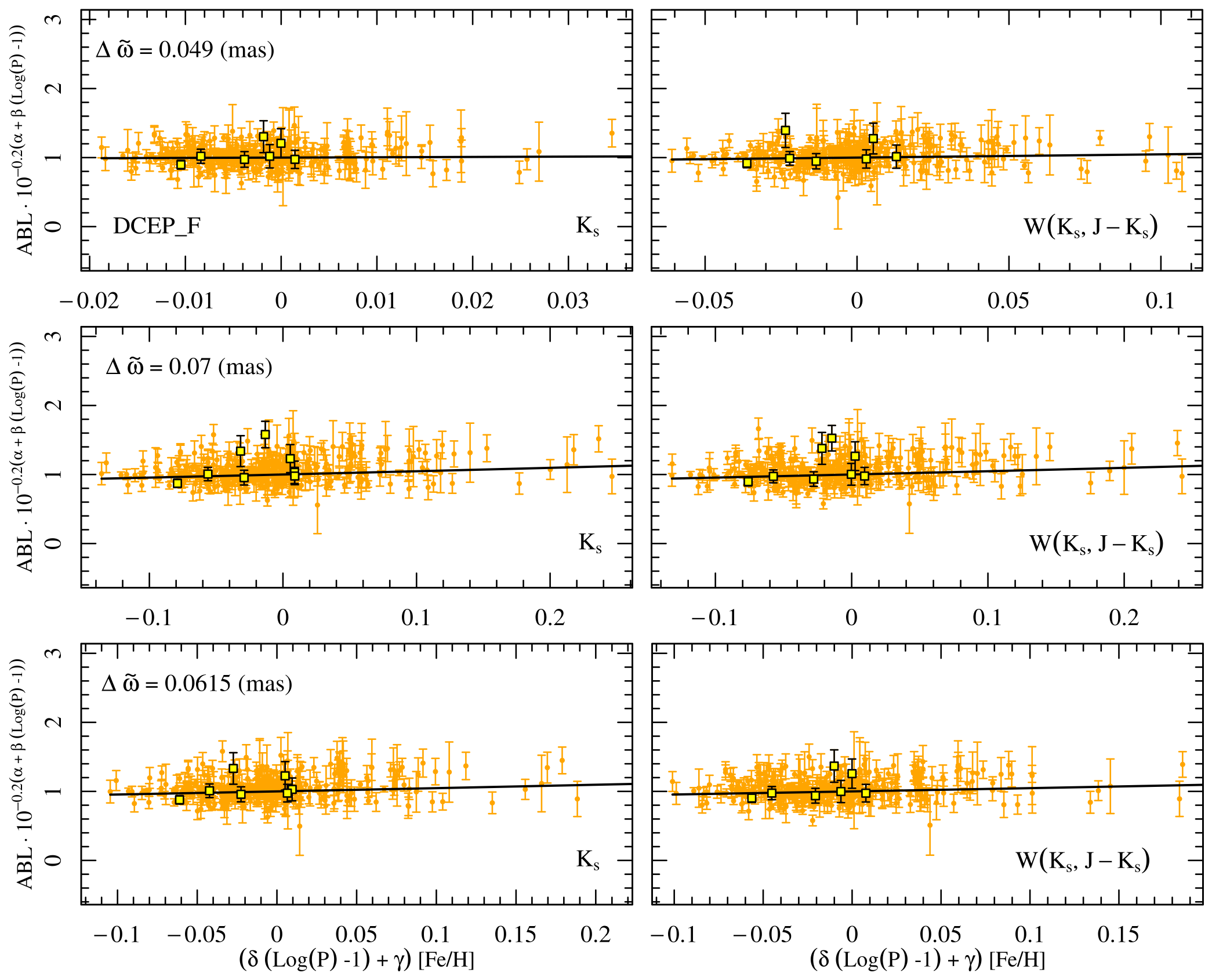}
   \caption{Same as in Fig.~\ref{fig:PW}, but fitting an ABL in the general form of Eq.~\ref{eqABLZbis1} (left panels) and  Eq.~\ref{eqABLZbis2} (right panels), for three different choices of the {\it Gaia} parallax zero point offset. The fit is presented only for fundamental mode pulsators as no sensible fit could be performed of the 10 DCEPs, due to their paucity in our  sample.} 
   \label{fig:ablCompleto}
    \end{figure*}

\begin{figure*}
   \centering
   \vbox{
   \includegraphics[width=12cm]{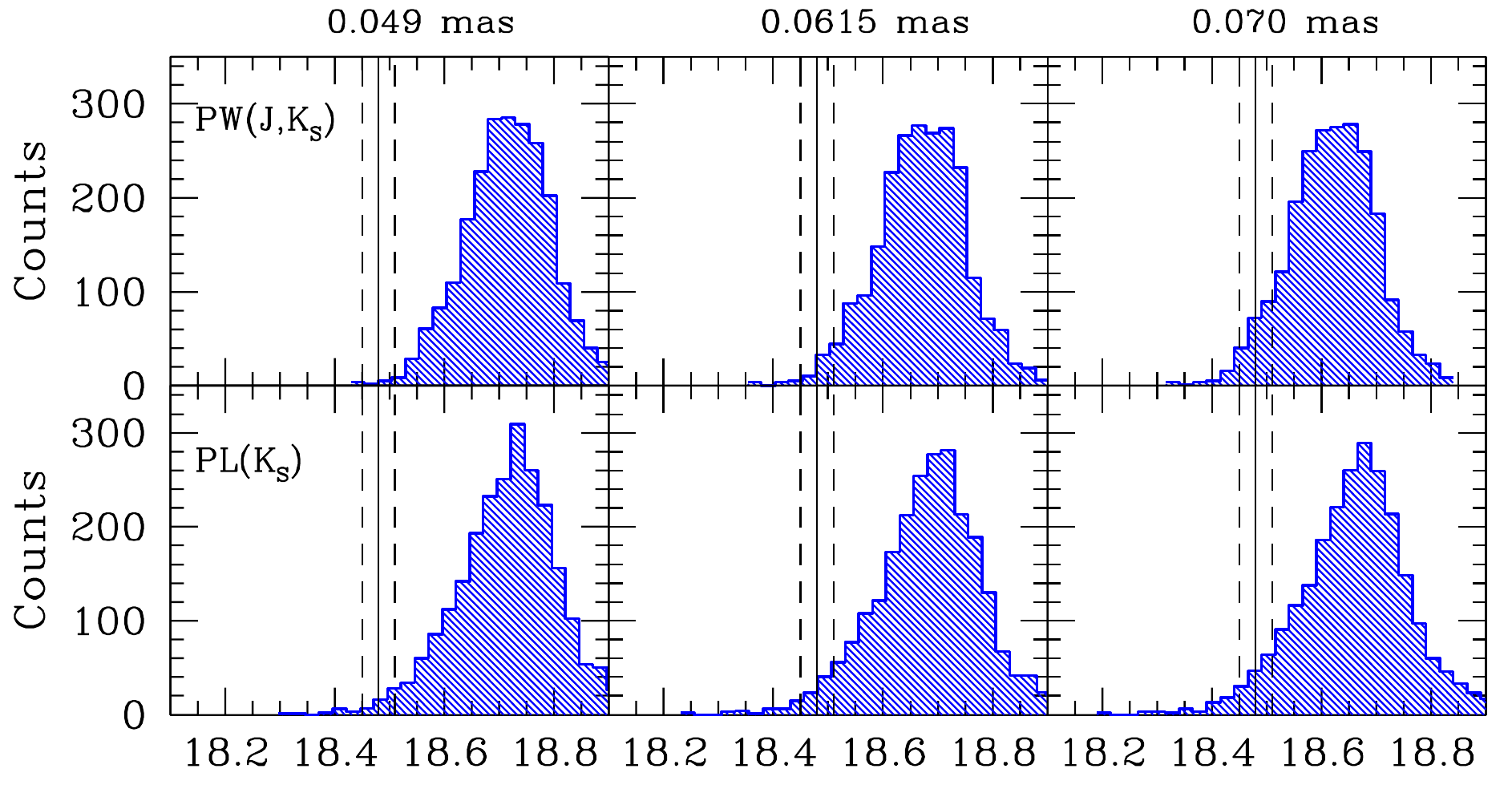}
   \includegraphics[width=12cm]{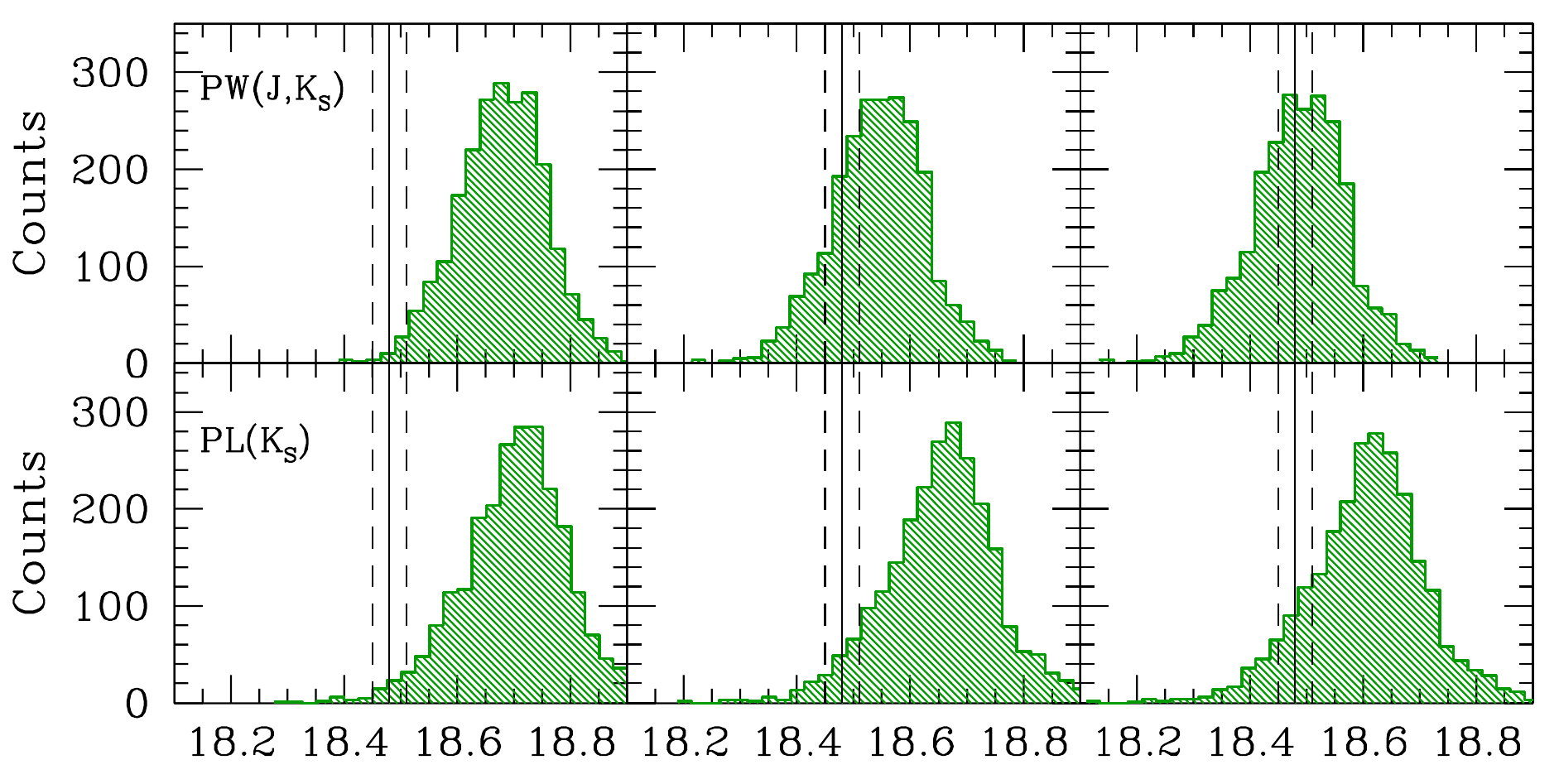}
   \includegraphics[width=12cm]{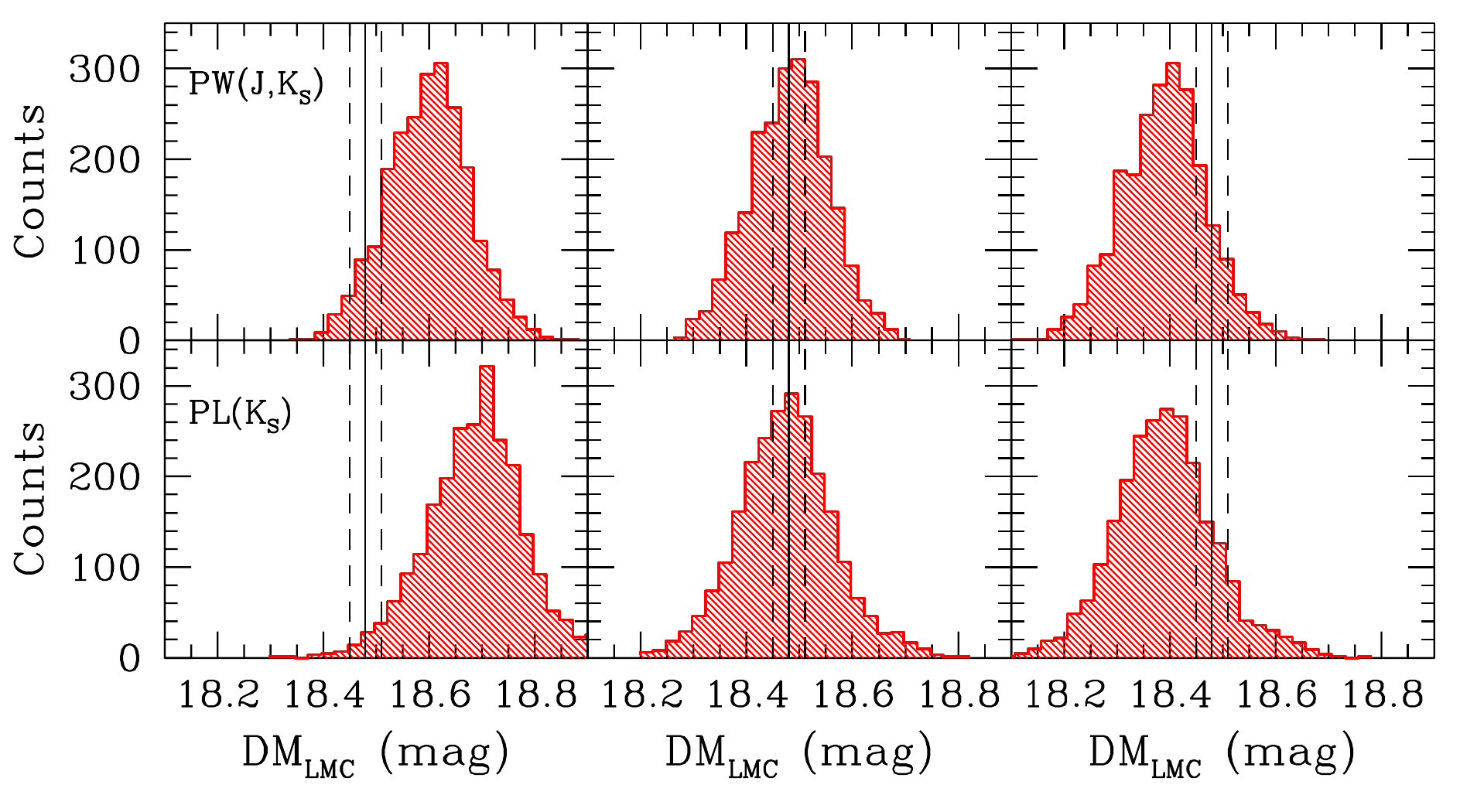}
}
   \caption{Distribution of the individual DMs for DCEP\_Fs in the LMC based on distances calculated using: i) the ABL in the form of Eq.~\ref{eqABL} and ~\ref{eqABL1} (top panels, blue histograms); ii) the ABL in the form of Eq.~\ref{eqABLZ1} and ~\ref{eqABLZ2} (middle panels, green histograms); iii) the ABL in the form of Eq.~\ref{eqABLZbis1} and ~\ref{eqABLZbis2} (bottom paneles, red histograms) and the NIR photometry of the VMC survey (see text for details). From left to right, the different panels show results obtained for three different values of the {\it Gaia} parallax zero point offset, namely,  $\Delta \varpi$=0.049, 0.0615 and 0.070 mas. A 
   vertical solid line shows the LMC distance modulus by \citet{Pietrzynski2019}, with dashed lines representing its 1 $\sigma$ uncertainty.} 
   \label{fig:histo}
    \end{figure*}

\section{Period-Luminosity-Metallicity and Period-Wesenheit-Metallicity relations in the NIR for the  Galactic DCEPs}\label{PL-PW}

We have used our final sample of 
373 DCEP\_Fs and 50 DCEP\_1Os to derive a new PL relation in the $K_S$ band and a new PW relation in the $J,K_S$ bands. First, we obtained dereddened $K_{S,0}$ magnitudes needed to construct
the PL$_{K_S}$ relation  by adopting  $A_{K_S}=0.34 E(B-V)$ \citep[ as derived by][using $R_V$=3.1]{Cardelli1989} where the $E(B-V)$ values are those collected in Paper I. The corresponding  Wesenheit magnitudes $w=Ks-0.69(J-K_S)$ are instead  reddening-free by construction. 

To avoid introducing any bias or sample truncation, we did not make any selection of the DCEPs based on their  parallax and/or  the ratio of parallax error over parallax, hence retaining also a few objects with negative parallax. To handle also   negative parallaxes we adopted the approach introduced by \citet{Feast1997} and 
\citet{Arenou1999},  and  computed the Astrometric Based Luminosity (ABL), {that in the simplest case is defined as follows:
\begin{eqnarray}
{\rm ABL}=10^{0.2 M_{K_{S,0}}}=10^{0.2[\alpha+\beta(\log P- \log P_0)]}=\varpi10^{0.2K_{S,0}-2} \label{eqABL} \\
{\rm ABL}=10^{0.2 W}=10^{0.2[\alpha+\beta(\log P- \log P_0)]}=\varpi10^{0.2w(J,K_S)-2} \label{eqABL1}
\end{eqnarray}
\noindent
where $K_{S,0}$,  $w(J,K_S)$,
the period P and the parallax $\varpi$ are the observables; the unknowns are the two parameters $\alpha$ and $\beta$ defining  the absolute quantity $M_{abs}=\alpha+\beta(\log P-\log P_0)$, which corresponds to the absolute magnitude $M_{K_{S,0}}$ in Eq.~\ref{eqABL} and to the absolute Wesenheit magnitude $W(J,K_S)$ in Eq.~\ref{eqABL1}. 
Note the use of a pivoting (logarithmic) period $P_0$ of 1.0 ($P_0$=10 d) for DCEP\_Fs and 0.3 ($P_0$= 2 d) for DCEP\_1Os, to reduce the correlation between the parameters of the fit. 
The $\alpha$, $\beta$ values were calculated by a weighted least-squares fit procedure with $\sigma$-clipping of the residuals,  adopting a double Median Absolute Deviation (MAD) with amplitude=4.5 MAD to limit the maximum number of rejected  objects to $\sim$10\%. Uncertainties were estimated through a bootstrap technique. The fit of the ABL function was carried out adopting three different values for the offset of the {\it Gaia}  parallax zero point, namely, $\Delta \varpi=0.049$ mas, following \citet{Groenewegen2018}, 
$\Delta \varpi=0.07$ mas according to \citet{Ripepi2019} and  the introduction of a third, intermediate, value: $\Delta \varpi=0.0615$ mas. 
The interested reader is referred to 
\citet{Groenewegen2018} and \citet{Ripepi2019}
 for a discussion of the {\it Gaia} parallax zero point offset for DCEPs \citep[see also][for a more general discussion of the topic]{Arenou2018,Lindegren2018}. 
The results of the fitting procedure are shown in Fig.~\ref{fig:PW0} where we have highlighted with different symbols the DCEPs analysed in Paper I.
The coefficients of the fit of  Eqs.~\ref{eqABL} and~\ref{eqABL1} are provided in the upper portion of Table~\ref{Tab:PW}. 
}
We first compare our results with the PL/PW relations defined by
the Large Magellanic Cloud (LMC) DCEPs, which are based on large samples, hence have very precise slopes (and intercepts). We use as a  reference the PL($K_S$) relations by \citet{Ripepi2012} and the  PW($J,K_S$) relations by \citet{Ripepi2020}, which are based on the NIR timeseries photometry collected by the Vista Magellanic Cloud Survey \citep[VMC][]{Cioni2011}\footnote{Similar results were obtained by other Authors \citep[e.g.][]{Inno2016}}. In particular, in the LMC these authors find PL and PW slopes of $\Delta K_S / \Delta \log P$=$-$3.295$\pm$0.018 and $\Delta w(J,K_S) / \Delta \log P$=$-$3.332$\pm$0.007 for the DCEP\_Fs, and   $\Delta K_S / \Delta \log P$=$-$3.471$\pm$0.035 and $\Delta w(J,K_S) / \Delta \log P$=$-$3.501$\pm$0.007 for the DCEP\_1Os. 
{A comparison with the values in the upper part of Table~\ref{Tab:PW} allows us to make the following  considerations: }
\begin{itemize} {
    \item notwithstanding an  increase by 12 units of the Galactic  DCEP\_1O sample provided by Paper I, their number is still too small to allow for precise results. Indeed, for any value of the parallax zero point offset, the slopes ($\beta$) of our MW PL and PW relations are significantly lower than for the LMC DCEP\_1Os and errors are very large; 
 
    \item the PL and PW relations of the MW DCEP\_Fs both show meaningful slopes with reasonably small errors. For all the values of $\Delta \varpi$ such slopes appear to be significantly lower than for the LMC DCEPs. The discrepancy is at the 3$\sigma$ level for the PL relation. 
    Since the average metallicity of the MW and LMC DCEPs differs by some $\sim0.3-0.4$ dex,
we can expect that  metallicity may   play a role in explaining these differences. 
To test this possibility, we added a metallicity term to the intercept of the PL and PW relations.  In this case, the ABL formulation becomes:}
\begin{eqnarray}
{\rm ABL}=10^{0.2 M_{K_{S,0}}}=10^{0.2\{\alpha+\beta(\log P- \log P_0)+\gamma{\rm [Fe/H]}\}}=\varpi10^{0.2K_{S,0}-2} \label{eqABLZ1} \\
{\rm ABL}=10^{0.2 W}=10^{0.2\{\alpha+\beta(\log P- \log P_0)+\gamma{\rm [Fe/H]}\}}=\varpi10^{0.2w(J,K_S)-2} \label{eqABLZ2}
\end{eqnarray}
\noindent
{We adopted the same procedure as above to fit Eqs.~\ref{eqABLZ1} and ~\ref{eqABLZ2} to the data, obtaining the coefficients listed in the middle portion of Table~\ref{Tab:PW}. Analysing these results we note that: i) for the DCEP\_Fs the metallicity term is negative and generally increasingly significant as $\Delta \varpi$ increases, especially for the PW relation; for the DCEP\_1Os  the results seem to show the same general trend (but the dependence is positive for the PL relation). However, 
the errors on the slopes and intercepts of the DCEP\_1O relations are up to one order of magnitude larger than for the  DCEP\_Fs; ii) for the  DCEP\_Fs, the slopes of both the PL and PW relations do not change significantly, decreasing slightly and thus becoming increasingly different than those of the LMC. The decrement is larger as the metallicity term increases, revealing an intricate inter-dependency between these parameters. In any case, it is difficult to escape the conclusion  that the slopes of the PL and PW relations for the MW DCEPs differ from those of the LMC DCEPs and seem to exhibit a non-negligible metallicity dependence. \\

  More in general, our results seem to indicate that there is a complicate interplay between the metallicity dependence, due to the higher average metallicity of the MW DCEPs with respect to the LMC DCEPs, the coefficients of the PL and PW relations and the zero point offset of the {\it Gaia} parallaxes. \\
In order to possibly shed light on this rather intricate scenario, we have adopted a more general form of the PL/PW relations which  takes into account a  metallicity dependence not only of the intercept, but also of the period coefficient.  This was rendered through the following general ABL functions:  }
\begin{eqnarray}
{\rm ABL}=10^{0.2(\alpha+(\beta+\delta {\rm [Fe/H])}(\log P- \log P_0) +\gamma {\rm [Fe/H]})}=\varpi10^{0.2K_{S,0}-2} \label{eqABLZbis1}    \\ 
{\rm ABL}=10^{0.2(\alpha+(\beta+\delta {\rm [Fe/H])}(\log P-\log P_0) +\gamma {\rm [Fe/H]})}=\varpi10^{0.2w(J,K_S)-2} \label{eqABLZbis2}  
\end{eqnarray}
\noindent
{which we proceeded to fit 
exactly in the same way as done in the previous cases, but considering only the DCEP\_F pulsators, as the DCEP\_1Os are too few in number to obtain meaningful results. The outcomes of this procedure are summarised in the bottom portion of  Table~\ref{Tab:PW} and shown in  Fig.~\ref{fig:ablCompleto} (top and medium panels). 
We note a general behaviour similar to the previous case, with the metallicity dependence on both intercept and slope increasing as $\Delta \varpi$ increases. In no case are 
the metallicity terms  significant for $\Delta \varpi$=0.049 mas, whereas the maximum significance is achieved for  $\Delta \varpi$=0.070 mas, when both the PL and PW relations show metallicity terms significant at 1-2$\sigma$ levels. 
In the intermediate case $\Delta \varpi$=0.0615 mas the metallicity dependence is significant at $\sim$1$\sigma$ level.
To summarise,  even if the uncertainties are still rather large, and except for the PL/PW relations with $\Delta \varpi$=0.049 mas, it was possible to find a meaningful metallicity dependence for both the slope (period coefficient) and intercept.  As far as we know, this is the first time that such a result is achieved, even though it still is more a qualitative than quantitative evidence.  We also note that the metallicity dependence of the slope appears to have the right sign, suggesting smaller slopes as [Fe/H] increases, as inferred from the comparison of LMC and  MW DCEPs and in remarkable agreement with the predictions of nonlinear convective pulsation models \citep[see e.g.][and references therein.]{Bono1999,Marconi2005}.
}

\subsection{Goodness of the fit}
{
In the previous part of Sect.~\ref{PL-PW} we have performed  several fits of the ABL function with an increasing number of coefficients to take into account the metallicity dependence, and for different values of $\Delta \varpi$, obtaining comparable values of the  ABL function residuals. Prompted by the referee, we now try to obtain an indication of the goodness of these fits. We choose to adopt the commonly used  Akaike Information Criterion \citep[AIC][]{Akaike2011}\footnote{We also calculated 
the Bayesian information criterion (BIC), which produced results very similar to the AIC, hence we only report and discuss the latter here.}. For this we first need to calculate the likelihood of each of the fitted ABL functions. Therefore, we define: 
\begin{equation}
    \ln{\mathcal{L}} = -0.5\sum_{i=1}^N \left [ \left (\frac{res_i}{s_i}\right )^2 + \ln (s_i^2)\right ]
\end{equation}
where $N$ is the number of fitted points, $res_i$ is the $i^{th}$ residual around the fitted relations ( Eqs. 1 to 6 above), while the  term $s_i$ is the sum in quadrature of the $i^{th}$ uncertainty  and the intrinsic scatter around the fitted relation. We have set the intrinsic scatter of the fitted relations equal to half the rms of the residuals, but checked that this choice does not affect the results significantly.
The values of the likelihood can now be used to estimate the AIC quantity, which is defined as:
\begin{equation}
    AIC=2k - 2\ln{\mathcal{L}}
\end{equation}
where k is the number of fitted parameters. The AIC values are reported in the last column of Tab.~\ref{Tab:PW}. By definition, the lower is the AIC value, the better is the fit. An analysis of Table~\ref{Tab:PW} reveals that introducing the metallicity terms on the intercept and slope has different effects on the PL and PW relations. Specifically, the introduction of a metallicity term on the intercept of the PL relation does not improve the fit to the data significantly, whereas the inclusion of a metallicity dependence for both  coefficients seems to improve the fit since, except for $\Delta \varpi$=0.049 mas, the coefficients of the PL relation become more  than $\sim 1\sigma$ significant.  
The introduction of the metallicity term on the intercept produces an improvement of the PW relation fit. The adoption of the metallicity term also on the slope, produces a slightly better or equal goodness of the fit. The best value of the AIC is obtained for the PW relation with $\Delta \varpi$=0.070 mas. 

In summary,  use of the AIC goodness of fit parameter allowed us to verify that the progressive inclusion of the metallicity term on the intercept and slope of the PL/PW relations, produces fits to the data which, except in the case of $\Delta \varpi$=0.049 mas, are  generally  (slightly) better than the cases when less parameters are used in the definition of the ABL function.
}

\section{Discussion and concluding remarks}
{
We can now test our PL/PW relations 
by applying them to the LMC DCEP\_F sample in  \citet{Ripepi2020} and 
comparing the 
resulting LMC  distance modulus  ($DM_{\rm LMC}$) 
to the distance to  the LMC  provided  by the geometric determination of \citet{Pietrzynski2019}: $DM_{\rm LMC}$=$18.48\pm0.03$ mag (including systematic errors),  which is currently considered one of the most accurate measure in the literature.

We calculate the absolute $M_{K_{0,S}}$ and Wesenheit magnitudes $W(J,K_S)$ for each LMC DCEP\_F in  \citet{Ripepi2020}, by inserting the proper periods in the respective equations 
in Table~\ref{Tab:PW} for the 
three different cases: i) with only the coefficients $\alpha$ and $\beta$; ii) with the addition of the $\gamma$ term; iii) with the further addition of the $\delta$ coefficient. 
In cases i) and ii) we are implicitly assuming that there is no  difference in the slope (period term) of the  PL/PW relations of the MW and LMC DCEPs. In  cases ii) and iii) we adopted an average metallicity for the LMC DCEPs of $\langle $[Fe/H]$\rangle$=$-$0.33 dex from \citet{Romaniello2008}. 
In all three cases, from the observed $K_{0,S}$ and $w(J,K_S)$ we obtained individual DM values for each LMC DCEP\_F. The distributions of these DMs are shown in Fig.~\ref{fig:histo} from top to bottom for the three cases i)-iii) and from left to the right for the three adopted values of $\Delta \varpi$=0.049, 0.0615 and 0.070 mas, respectively. In each panel we compared the DM distributions with the LMC geometric distance by \citet{Pietrzynski2019}. In all cases the  histograms show rather narrow distributions with average dispersion   on the order of 0.08-0.09 mag. 

The top  panels of Fig.~\ref{fig:histo} (blue histograms)  show that in all cases,  solution i), no metallicity dependence,  provides $<DM_{\rm LMC}>$ values which are much larger than the LMC geometric distance by \citet{Pietrzynski2019}. A significantly  larger $\Delta \varpi$ offset would be needed to reconcile the two results. 
The middle panels of Fig.~\ref{fig:histo} (green histograms)  corresponding to case ii) (metallicity dependence only on the intercept), still provide too large $<DM_{\rm LMC}>$ values except in the case of the PW relation with $\Delta \varpi$=0.070 mas, which,  however, has the largest difference in slope with respect to the observed LMC PW relation. 
Finally,  the bottom panels of Fig.~\ref{fig:histo} (red histograms), corresponding  to  case iii) (metallicity dependence of  both slope and intercept) are in very good agreement with the $<DM_{\rm LMC}>$ value by \citet{Pietrzynski2019} for $\Delta \varpi$=0.0615 mas, whereas the other two values of $\Delta \varpi$ produce either lower or larger $<DM_{\rm LMC}>$ estimates. 

Therefore, the general formulation of the PL/PW relations, is not only able to qualitatively explain the different slopes of the PL/PW relations in the MW and LMC, but it also allows us to obtain an  estimate of the zero point offset of the {\it Gaia} DR2 parallaxes for DCEPs. An error on the offset  value can be inferred from the uncertainty on the LMC distance modulus by \citet{Pietrzynski2019}. This corresponds to  0.03 mag, which 
 translates into a 0.004 mas error in the parallax offset. Therefore, our best estimate for the {\it Gaia} DR2 zero point parallax offset for our DCEP dataset is 0.0615$\pm$0.004 mas. It is important to remark that this zero point offset has to be considered as an average valid only for the particular DCEP dataset used here, as in general, the zero point offset of the {\it Gaia} parallaxes varies for different groups of objects and different positions in the sky \citep[see e.g.][]{Arenou2018,Lindegren2018,Leung2019}. 

To conclude, the general PL/PW relations provided in the bottom part of Table~\ref{Tab:PW} could represent a step forward in our ability to 
measure distances through  DCEPs, with 
a potential great impact on the cosmic distance scale and the estimate of H$_0$. 
Additional spectroscopic measurements for DCEPs both in the MW and the LMC, as well as more precise parallaxes expected from the forecoming {\it Gaia} Early Data Release 3 will be fundamental to confirm the results presened in this paper.
}

\end{itemize}

\begin{acknowledgements}

{We wish to thank the  anonymous Referee for his/her  suggestions, which helped to improve the paper.}
This work has made use of data from the European Space Agency (ESA) mission
{\it Gaia} (\url{https://www.cosmos.esa.int/gaia}), processed by the {\it Gaia} Data Processing and Analysis Consortium (DPAC,
\url{https://www.cosmos.esa.int/web/gaia/dpac/consortium}). Funding for the DPAC has been provided by national institutions, in particular the institutions participating in the {\it Gaia} Multilateral Agreement.
In particular, the Italian participation
in DPAC has been supported by Istituto Nazionale di Astrofisica
(INAF) and the Agenzia Spaziale Italiana (ASI) through grants I/037/08/0,
I/058/10/0, 2014-025-R.0, and 2014-025-R.1.2015 to INAF (PI M.G. Lattanzi).
V.R., M.M. and G.C. acknowledge partial support from the project "MITiC: MIning The Cosmos Big Data and Innovative Italian Technology for Frontier Astrophysics and Cosmology”  (PI B. Garilli).
RM thanks his wife and her parents that made possible his collaboration to this work during the current pandemic period.
\end{acknowledgements}

%

\begin{thebibliography}{}

\bibitem[Akaike(2011)]{Akaike2011} Akaike, H.\ 2011, Akaike’s Information Criterion. In: Lovric M. (eds) International Encyclopedia of Statistical Science. Springer, Berlin, Heidelberg

\bibitem[Arenou \& Luri(1999)]{Arenou1999} Arenou, F., \& Luri, X.\ 1999, Harmonizing Cosmic Distance Scales in a Post-HIPPARCOS Era, 167, 13 

\bibitem[Arenou et al.(2018)]{Arenou2018} Arenou, F., Luri, X., Babusiaux, C., et al.\ 2018, \aap, 616, A17


\bibitem[Bono et al.(1999)]{Bono1999} Bono, G., Caputo, F., Castellani, V., et al.\ 1999, \apj, 512, 711

\bibitem[Bono et al.(2010)]{Bono2010} Bono, G., Caputo, F., Marconi, M., et al.\ 2010, \apj, 715, 277


\bibitem[Cardelli et al.(1989)]{Cardelli1989} Cardelli, J.~A., Clayton, G.~C., \& Mathis, J.~S.\ 1989, \apj, 345, 245


\bibitem[Catanzaro et al.(2020)]{Catanzaro2020} Catanzaro, G., Ripepi, V., Clementini, G., et al.\ 2020, A\&A, in press doi.org/10.1051/0004-6361/202038486

\bibitem[Chen et al.(2018)]{Chen2018} Chen, X., Wang, S., Deng, L., et al.\ 2018, \apjs, 237, 28

\bibitem[Cioni et al.(2011)]{Cioni2011} Cioni, M.-R.~L., Clementini, G., Girardi, L., et al.\ 2011, \aap, 527, A116


\bibitem[Clementini et al.(2019)]{Clementini2019} Clementini, G., Ripepi, V., Molinaro, R., et al.\ 2019, \aap, 622, A60


\bibitem[Di Criscienzo et al.(2013)]{Dicriscienzo2013} Di Criscienzo, M., Marconi, M., Musella, I., Cignoni, M., \& Ripepi, V.\ 2013, \mnras, 428, 212 

\bibitem[Fausnaugh et al.(2015)]{Fausnaugh2015} Fausnaugh, M.~M., Kochanek, C.~S., Gerke, J.~R., et al.\ 2015, \mnras, 450, 3597

\bibitem[Feast \& Catchpole(1997)]{Feast1997} Feast, M.~W. \& Catchpole, R.~M.\  1997, \mnras, 286, L1

\bibitem[Fiorentino et al.(2007)]{Fiorentino2007} Fiorentino, G., Marconi, M., Musella, I., \& Caputo, F.\ 2007, \aap, 476, 863 

\bibitem[Fiorentino et al.(2013)]{Fiorentino2013} Fiorentino, G., Musella, I., \& Marconi, M.\ 2013, \mnras, 434, 2866 

\bibitem[Freedman, \& Madore(2011)]{Freedman2011} Freedman, W.~L., \& Madore, B.~F.\ 2011, \apj, 734, 46

\bibitem[Freedman et al.(2012)]{Freedman2012} Freedman, W.~L., Madore, B.~F., Scowcroft, V., et al.\ 2012, \apj, 758, 24

\bibitem[Gaia Collaboration et al.(2017)]{Clementini2017} Gaia Collaboration, Clementini, G., Eyer, L., et al.\ 2017, \aap, 605, A79

\bibitem[Gaia Collaboration et al.(2016)]{Gaia2016} Gaia Collaboration, Prusti, T., de Bruijne, J.~H.~J., et al.\ 2016, \aap, 595, A1


\bibitem[Gieren et al.(2018)]{Gieren2018} Gieren, W., Storm, J., Konorski, P., et al.\ 2018, \aap, 620, A99 


\bibitem[Groenewegen(2013)]{Groenewegen2013} Groenewegen, M.~A.~T.\ 2013, \aap, 550, A70


\bibitem[Groenewegen(2018)]{Groenewegen2018} Groenewegen, M.~A.~T.\ 2018, \aap, 619, A8


\bibitem[Huang et al.(2020)]{Huang2020} Huang, C.~D., Riess, A.~G., Yuan, W., et al.\ 2020, \apj, 889, 5

\bibitem[Inno et al.(2016)]{Inno2016} Inno, L., Bono, G., Matsunaga, N., et al.\ 2016, \apj, 832, 176

\bibitem[Jayasinghe et al.(2018)]{Jayasinghe2018} Jayasinghe, T., Kochanek, C.~S., Stanek, K.~Z., et al.\ 2018, \mnras, 477, 3145

\bibitem[Kodric et al.(2013)]{Kodric2013} Kodric, M., Riffeser, A., Hopp, U., et al.\ 2013, \aj, 145, 106


\bibitem[Leavitt \& Pickering(1912)]{Leavitt1912} Leavitt, H.~S., \& Pickering, E.~C.\ 1912, Harvard College Observatory Circular, 173, 1 

\bibitem[Leung \& Bovy(2019)]{Leung2019} Leung, H.~W. \& Bovy, J.\ 2019, \mnras, 489, 2079


\bibitem[Lindegren et al.(2018)]{Lindegren2018} Lindegren, L., Hern{\'a}ndez, J., Bombrun, A., et al.\ 2018, \aap, 616, A2



\bibitem[Macri et al.(2006)]{Macri2006} Macri, L.~M., Stanek, K.~Z., Bersier, D., et al.\ 2006, \apj, 652, 1133


\bibitem[Madore(1982)]{Madore1982} Madore, B.~F.\ 1982, ApJ, 253, 575 

\bibitem[Marconi et al.(2005)]{Marconi2005} Marconi, M., Musella, I., \& Fiorentino, G.\ 2005, \apj, 632, 590

\bibitem[Ngeow et al.(2012b)]{Ngeow2012b} Ngeow, C.-C., Kanbur, S.~M., Bellinger, E.~P., et al.\ 2012b, \apss, 341, 105 

\bibitem[Pejcha, \& Kochanek(2012)]{Pejcha2012} Pejcha, O., \& Kochanek, C.~S.\ 2012, \apj, 748, 107


\bibitem[Pietrzy{\'n}ski et al.(2019)]{Pietrzynski2019} Pietrzy{\'n}ski, G., Graczyk, D., Gallenne, A., et al.\ 2019, \nat, 567, 200

\bibitem[Planck Collaboration et al.(2018)]{Planck2018} Planck Collaboration, Aghanim, N., Akrami, Y., et al.\ 2018, arXiv e-prints, arXiv:1807.06209


\bibitem[Reid et al.(2019)]{Reid2019} Reid, M.~J., Pesce, D.~W., \& Riess, A.~G.\ 2019, \apjl, 886, L27

\bibitem[Riess et al.(2016)]{Riess2016} Riess, A.~G., Macri, L.~M., Hoffmann, S.~L., et al.\ 2016, \apj, 826, 56

\bibitem[Riess et al.(2018)]{Riess2018} Riess, A.~G., Casertano, S., Yuan, W., et al.\ 2018b, \apj, 861, 126

\bibitem[Riess et al.(2019)]{Riess2019} Riess, A.~G., Casertano, S., Yuan, W., et al.\ 2019, \apj, 876, 85

\bibitem[Ripepi et al.(2012)]{Ripepi2012} Ripepi, V., Moretti, M.~I., Marconi, M., et al.\ 2012, \mnras, 424, 1807

\bibitem[Ripepi et al.(2019)]{Ripepi2019} Ripepi, V., Molinaro, R., Musella, I., et al.\ 2019, \aap, 625, A14

\bibitem[Ripepi et al.(2020)]{Ripepi2020} Ripepi, V., Molinaro, R., Marconi, M., et al.\ 2020, arXiv e-prints, arXiv:2002.10584

\bibitem[Romaniello et al.(2008)]{Romaniello2008} Romaniello, M., Primas, F., Mottini, M., et al.\ 2008, \aap, 488, 731

\bibitem[Sandage et al.(2006)]{Sandage2006} Sandage, A., Tammann, G.~A., Saha, A., et al.\ 2006, \apj, 653, 843

\bibitem[Shappee, \& Stanek(2011)]{Shappee2011} Shappee, B.~J., \& Stanek, K.~Z.\ 2011, \apj, 733, 124

\bibitem[Udalski et al.(2018)]{Udalski2018} Udalski, A., Soszy{\'n}ski, I., Pietrukowicz, P., et al.\ 2018, \actaa, 68, 315


\bibitem[Yuan et al.(2019)]{Yuan2019} Yuan, W., Riess, A.~G., Macri, L.~M., et al.\ 2019, \apj, 886, 61

\bibitem[Wong et al.(2020)]{Wong2020} Wong, K.~C., Suyu, S.~H., Chen, G.~C.-F., et al.\ 2020, \mnras, doi:10.1093/mnras/stz3094



\end{thebibliography}
%

%


\end{document}